\documentclass{article}

\usepackage{amsmath}
\usepackage{amsfonts}
\usepackage{amssymb}
\usepackage{amsthm}
\usepackage{newlfont}
\usepackage{graphicx}

\makeatletter
\@addtoreset{equation}{section}
\def\theequation{\arabic{section}.\arabic{equation}}
\makeatother

\newcommand{\KT}{{\text{KT}}}
\renewcommand{\r}{{\mathbf{r}}}
\newcommand{\F}[2]{\mathop{{}_#1F_#2}}

\newcommand{\eqat}[1]{\mathop{\underset{#1}{=}}}
\newcommand{\simat}[1]{\mathop{\underset{#1}{\sim}}}

\newcommand{\I}{{\cal I}}
\newlength{\GraphicsWidth}
\begin{document}

\title{Equation of state in the fugacity format for the two-dimensional
  Coulomb gas}
\date{}
\author{Gabriel T\'ellez\footnote{email: gtellez@uniandes.edu.co}\\
Departamento de F\'\i sica, Universidad de Los Andes\\
A.~A.~4976 Bogot\'a, Colombia.
}

\maketitle

\begin{abstract}
We derive the exact general form of the equation of state, in the
fugacity format, for the two-dimensional Coulomb gas. Our results are
valid in the conducting phase of the Coulomb gas, for temperatures
above the Kosterlitz-Thouless transition. The derivation of the
equation of state is based on the knowledge of the general form of the
short-distance expansion of the correlation functions of the Coulomb
gas. We explicitly compute the expansion up to order $O(\zeta^6)$ in
the activity $\zeta$. Our results are in very good agreement with
Monte Carlo simulations at very low density.
\end{abstract}


\section{Introduction and summary of results}

The system under consideration is a classical two component Coulomb
gas composed of positive and negative particles with charges $+1$ and
$-1$. The particles live in a two dimensional plane and they are small
impenetrable disks of diameter $\sigma$. The interaction between two
charges $q$ and $q'$ at a distance $r$ from each other is
\begin{equation}
  \label{eq:Coulomb-pot}
  v(r)=
  \begin{cases}
    -q q' \ln \frac{r}{L} & r>\sigma\\
    +\infty & r \leq \sigma
  \end{cases}
\end{equation}
where $L$ is an arbitrary length scale fixing the zero of the
potential. This is the two dimensional version of the restricted
primitive model for electrolytes. We shall work using the grand
canonical formalism with fugacity $\lambda$ (dimensions $length^{-2}$)
and inverse reduced temperature (coulombic coupling) $\beta$. The
arbitrary length scale $L$ can be absorbed in the fugacity by defining
the rescaled fugacity $z=\lambda L^{\beta/2}$ which has dimensions
$length^{(\beta-4)/2}$. A dimensionless activity which will prove
useful later can be defined as $\zeta=z \sigma^{(4-\beta)/2}=\lambda
\sigma^2 (L/\sigma)^{\beta/2}$. Notice that if $L$ is chosen as
$L=\sigma$, $\zeta=\lambda \sigma^2$ does not depend on $\beta$ for
fixed fugacity $\lambda$. In the thermodynamic limit only neutral
configurations are relevant. Let $n_{+}=n_{-}$ be the density of
positive (negative) particles. The total number density is
$n=n_{+}+n_{-}=2n_{+}$.

In the low density limit $n\sigma^2\to0$, there are two values of the
coupling $\beta$ of special interest. At $\beta=\beta_{\KT}=4$ the
system undergoes the Kosterlitz-Thouless transition of infinite
order~\cite{KT}. In the high temperature phase $\beta<\beta_{\KT}$,
the system is in a conducting phase with free ions that can screen
external charges. The correlations have an exponential decay and they
satisfy several screening sum rules, for instance the
Stillinger-Lovett sum rule~\cite{SL}. In the low temperature phase,
for $\beta>\beta_{\KT}$, the gas is in a dielectric phase where all
charges are bound forming dipolar pairs. The perfect screening sum
rule is no longer satisfied.

The other value for $\beta$ of interest is $\beta=2$. For $\beta<2$
the thermodynamic quantities and correlation functions of the system
have a finite value in the limit of point particles $\sigma=0$, while
for $2\leq\beta<4$ at fixed fugacity $z$, the density, the free energy
and internal energy of the system diverge when $\sigma\to0$. This is
due to the collapse of pairs of point particles of opposite sign. On
the other hand, it is believed~\cite{KS} that the truncated density
correlation functions remain finite in the limit $\sigma\to 0$ when
$2\leq\beta<4$.

For $\beta<2$ and $\sigma=0$, the equation of state for the pressure
$p$ of the plasma has been known for a long time~\cite{SP}. A simple
scaling argument gives the volume dependence of the free energy which
leads to the pressure
\begin{equation}
  \beta p = \left(1-\frac{\beta}{4}\right) n\,.
\end{equation}

On the other hand the temperature dependence of the free energy is
highly non trivial. Only recently, exact results for the full
thermodynamics of the two-dimensional Coulomb gas, in the region
$\beta<2$ and $\sigma=0$, have been obtained by \v{S}amaj and
Trav\v{e}nec~\cite{ST}. These results have been obtained using the
equivalence between the classical Coulomb gas and the quantum
sine-Gordon model. In two dimensions this model is integrable, the
free energy is known in terms of the soliton mass~\cite{DV}, and the
relation between the soliton mass and the coupling of the sine-Gordon
model (i.~e.~the fugacity of the Coulomb gas) in the conformal
normalization has been found~\cite{Zamolo-mass-scale}. This gives the
exact density--fugacity relationship for the Coulomb gas, which allows
to find all the thermodynamic quantities of the system~\cite{ST}.

In the region $2\leq\beta<4$, since the density diverges in the limit
$n\sigma^2\to0$ for fixed fugacity, it is more appropriate to study
the fugacity expansion of the pressure, rather than its density
expansion.  In Ref.~\cite{GN}, Gallavotti and Nicol\'o considered a
version of the Coulomb gas with a soft short-distance cutoff. They
proved that a Mayer series expansion of the pressure in integer powers
of the fugacity have well defined coefficients up to order $2l$ for
$\beta>\beta_{l}$, where
\begin{equation}
  \beta_{l}=4\left(1-\frac{1}{2l}\right)\,\qquad l=1, 2, 3, \ldots
\end{equation}
whereas higher order Mayer coefficients diverge. For $\beta>4$ all
Mayer coefficients are finite, while for $\beta<2$ all Mayer
coefficients diverge.

Their findings lead them to conjecture that the plasma undergoes a
series of intermediate phase transitions at $\beta=\beta_l$ from the
conducting phase at $\beta=\beta_1=2$ up to the dielectric phase at
$\beta=\beta_{\infty}=\beta_{\KT}=4$, as opposed to the traditional
Kosterlitz-Thouless scenario where the conducting-dielectric phase
transition takes place at $\beta=4$.

Fisher \textit{et al.}~\cite{FLL} denied this conjecture. They proposed
an ansatz for the pressure, which, in our notations
($\zeta=z\sigma^{(4-\beta)/2}$), reads
\begin{equation}
  \label{eq:Fisher}
  \beta p=b_{\psi}(\beta) z^{4/(4-\beta)}\left[
    1+e(\beta,\zeta)\right] +\frac{1}{\sigma^2} \sum_{l=1}^{\infty}
    \bar{b}_{2l}(\beta) \zeta^{2l} \,.
\end{equation}
In this ansatz, they conjectured that $b_\psi(\beta)$ and
$\bar{b}_{2l}(\beta)$ are analytic for $\beta<4$ and that
$e(\beta,\zeta)$ is an analytic function of $\beta$ for $\beta<4$ and
is also analytic in $\zeta$ for $\zeta>0$. Furthermore, for
$n\sigma^2\to0$, at fixed $z$, $e(\beta, z\sigma^{(4-\beta)/2})\to0$.

While the ansatz~(\ref{eq:Fisher}) is fully compatible with Gallavotti
and Nicol\'o findings, these later conditions on $b_\psi(\beta)$,
$\bar{b}_{2l}(\beta)$, and $e(\beta,\zeta)$ imply that the pressure
exhibits no singularities up to $\beta=4$, thus there are no
intermediate phase transitions.

Using the exacts results for $\beta<2$ and $\sigma=0$~\cite{ST},
Kalinay and \v{S}amaj~\cite{KS} devised a method to obtain results for
the thermodynamic properties of the Coulomb gas in the low density
limit $n\sigma^2\ll 1$ up to $\beta<3$. Their findings confirm the
form~(\ref{eq:Fisher}) of the ansatz proposed by Fisher \textit{et
al.}~but the analytic structure of the coefficients $b_\psi(\beta)$
and $\bar{b}_{2l}(\beta)$ is different. They have simple poles at
$\beta=\beta_l$ but they conjectured that a cancellation occurs. At
$\beta=\beta_{l}$, the exponent of the fugacity in the nonanalytic
part of $\beta p$ is integer: $4/(4-\beta_l)=2l$. Then it turns out
that residues of $\bar{b}_{2l}(\beta)$ and $b_\psi(\beta)$ at
$\beta=\beta_l$ are opposite, thus giving no singularities for the
pressure at $\beta=\beta_l$, confirming the absence of intermediate
phase transitions.

The cancellation of singularities was verified in Ref.~\cite{KS} at
the first threshold $\beta=\beta_1=2$, and conjectured for the other
thresholds. The aim of this work is to extend further the analysis of
Ref.~\cite{KS}. One important ingredient in the analysis of
Ref.~\cite{KS} is the knowledge of the short-distance expansion of the
density correlations functions of the Coulomb gas, $n_{+-}^{(2)}(r)$
and $n_{++}^{(2)}(r)$. The cancellation at $\beta=2$ was obtained in
Ref.~\cite{KS} using the fact that, at the lowest order when $r\to0$,
$n_{+-}^{(2)}(r)-n_{++}^{(2)}(r) \sim z^2 r^{-\beta}$. 

In a recent work~\cite{GT-small-r}, we presented the general framework
to obtain higher order terms of this expansion and explicitly computed
the two next order terms of the short-distance expansion of the
correlation functions. Based on this previous
analysis~\cite{GT-small-r}, we will show that the function
$e(\beta,\zeta)$ in the ansatz~(\ref{eq:Fisher}) is nonanalytic in
$\zeta$. Actually, we will show that the ansatz~(\ref{eq:Fisher})
should be generalized to
\begin{equation}
  \label{eq:ansatzGT}
  \beta p=\frac{1}{\sigma^2}\sum_{l=1}^{\infty} \bar{b}_{2l}(\beta)
  \zeta^{2l} +b_{\psi}(\beta) z^{\frac{4}{4-\beta}}\left[
  1+e_{1,0}(\beta,\zeta)\right]
  +\frac{1}{\sigma^2}\sum_{m=0}^{\infty}\sum_{n}\nolimits^{*}
  \zeta^{\frac{4m+\beta n^2}{4-\beta}} \tilde{e}_{n,m}(\beta,\zeta)
  \,.
\end{equation}
The sum $\sum_{n}\nolimits^{*}$ is for $n\geq 2$ when $m=0$ and for
$n\geq 0$ for $m\geq 1$. For $m=1$ the function
$e_{n,1}(\beta,\zeta)=0$.  The functions $\tilde{e}_{n,m}$ and
$e_{1,0}$ admit an expansion in integer powers of $\zeta$ and
$\ln\zeta$,
\begin{equation}
  \label{eq:tilde-e}
  \tilde{e}_{n,m}(\beta,\zeta)=
  \sum_{k=0}^{\infty} \tilde{e}_{n,m,k}(\beta,\ln \zeta) \,\zeta^{2k^{*}+n}
\end{equation}
\begin{equation}
  \label{eq:e}
  e_{1,0}(\beta,\zeta)= \sum_{k=0}^{\infty}
  e_{1,0,k}(\beta,\ln \zeta) \,\zeta^{2k+2}
\end{equation}
where $k^{*}=k+1$ for $n=0, 1$ and $k^{*}=k$ for $n\geq 2$. Notice
that the lowest order in this expansion is at least $\zeta^2$. Thus,
in the limit $n\sigma^2\to 0$, fixed $z$, and $\beta<4$, these terms
are irrelevant in the sense that $\tilde{e}_{n,m}(\beta,
z\sigma^{(4-\beta)/2})\to 0$.

The explicit calculation of $b_{\psi}(\beta)$ and $\bar{b}_{2}(\beta)$
was done in Ref.~\cite{KS}. Here we will compute explicitly
$\bar{b}_{4}(\beta)$ and $e_{1,0,0}(\beta,\ln \zeta)$. We will show
that the cancellation mechanism between $b_\psi(\beta)$ and
$\bar{b}_{2l}(\beta)$ conjectured in Ref.~\cite{KS} indeed takes place
for $l=2$ at $\beta=\beta_2=3$.

The outline of this paper is the following. In
section~\ref{sec:CG-SG}, we will recall some basic facts about the
exact results~\cite{ST} for $\beta<2$ and $\sigma=0$ and about the
general strategy proposed by \v{S}amaj and Kalinay~\cite{KS} to obtain
the thermodynamics of the Coulomb gas for $\beta>2$ in low density
limit. In section~\ref{sec:pressure-general}, we prove that the
equation of state in the fugacity format has the form proposed in
Eq.~(\ref{eq:ansatzGT}). In section~\ref{sec:pressure-coefs}, we
compute explicitly the coefficients $\bar{b}_{4}(\beta)$ and
$e_{1,0,0}(\beta,\ln \zeta)$ and verify the cancellation mechanism
between $b_\psi(\beta)$ and $\bar{b}_{4}(\beta)$ at
$\beta=\beta_2=3$. In section~\ref{sec:mc}, from our analytical
results for the equation of state, we compute the internal energy and
the specific heat of the two-dimensional Coulomb gas and we compare
our results with ones obtained from Monte Carlo
simulations~\cite{Caillol-Levesque}.


\section{Previous results and general strategy}
\label{sec:CG-SG}


\subsection{The Coulomb gas of point charges for $\beta<2$}

For point particles, $\sigma=0$ and $\beta<2$, the classical Coulomb
gas can be mapped into the Euclidean quantum sine-Gordon model by
carrying out a Hubbard-Stratonovich transformation. The grand
canonical partition function $\Xi$ of the Coulomb gas can be written
as
\begin{equation}
  \Xi=\frac{\int {\cal D}\phi \exp[-S(z)]}{\int {\cal D}\phi \exp[-S(0)]}
\end{equation}
with the sine-Gordon action
\begin{equation}
  \label{eq:sG-action}
  S(z)=-\int d^2\r \left[
    \frac{1}{16\pi}\phi(\r)\Delta\phi(\r)+2z\cos(b\phi(\r))
    \right]
\end{equation}
where we defined
\begin{equation}
  b^2=\beta/4
  \,.
\end{equation}
Under this mapping, the bulk density and two-body densities of charges
$q=\pm1$ and $q'=\pm1$ are given
by~\cite{ST,Samaj-Janco-TCP-large-distance-correl}
\begin{equation}
  \label{eq:bulk-density}
  n_q=z \langle e^{ibq\phi} \rangle
\end{equation}
and
\begin{equation}
  \label{eq:density-correl}
  n_{qq'}^{(2)}(|\r-\r'|)=z^2
  \langle e^{ibq\phi(\r)} e^{ibq'\phi(\r')} \rangle
\end{equation}
where the averages are taken with respect to the sine-Gordon
action~(\ref{eq:sG-action}). 

The complete mapping between the classical Coulomb gas and the
sine-Gordon model requires~\cite{ST,Samaj-tcp-review} to use the
conformal normalization~\cite{Zamolo-mass-scale}, where, when $z\to0$,
the free fields are normalized according to $\langle e^{ib\phi(0)}
e^{-ib\phi(\r)} \rangle_{z=0}=r^{-\beta}$. Under this conformal
normalization, the expectation value of exponential fields is
known~\cite{Lukyanov-Zamolod-exp-field, Fateev-reflexion}
\begin{equation}
  \label{eq:eibQ}
  \langle e^{i b Q \phi} \rangle = \left(\frac{\pi
  z}{\gamma(\beta/4)}\right)^{\frac{\beta Q^2}{4-\beta}} \exp[I_b(Q)]
\end{equation}
with $\gamma(x)=\Gamma(x)/\Gamma(1-x)$ where $\Gamma(x)$ is the Euler
Gamma function, and
\begin{equation}
  \label{eq:IbQ}
  I_b(Q)=\int_0^\infty
  \frac{dt}{t}
  \left[
    \frac{\sinh^2(2Qb^2 t)}{2\sinh(b^2 t)\sinh(t)\cosh[(1-b^2)t]}
    -2Q^2 b^2 e^{-2t}
    \right]
  \,.
\end{equation}
This expression is valid for $\beta|Q|<2$, otherwise the integral
$I_b(Q)$ diverges, but it is possible to do an analytic
continuation~\cite{Samaj-hard-core} of this formula for other values
of $\beta Q$ using a reflexion formula satisfied by $\langle
e^{ibQ\phi} \rangle$ presented in Ref.~\cite{Fateev-reflexion}.  For
$Q=\pm1$, the integral~(\ref{eq:IbQ}) can be computed explicitly and
\begin{equation}
  \label{eq:eibphi}
  \langle e^{i b \phi} \rangle =\langle e^{-i b \phi} \rangle =
  2\left(\frac{\pi z}{\gamma(\beta/4)}\right)^{\frac{\beta}{4-\beta}}
  \left(\frac{\Gamma(\xi/2)}{\Gamma(\frac{1+\xi}{2})}\right)^2
  \frac{\tan(\pi\xi/2)}{(4-\beta)\gamma(\beta/4)}
\end{equation}
where $\xi=\beta/(4-\beta)$.

Equation~(\ref{eq:bulk-density}) combined with~(\ref{eq:eibphi}) gives
the exact density--fugacity relationship~\cite{ST}
\begin{equation}
  \label{eq:density-sigma0}
  n=n[z,0]= z^{4/(4-\beta)} \, \frac{4}{4-\beta}\, b_{\psi}(\beta)
\end{equation}
with
\begin{equation}
  \label{eq:bpsi}
  b_{\psi}(\beta)=
  \frac{\pi^{\frac{\beta}{4-\beta}}}{
    (\gamma(\beta/4))^{\frac{4}{4-\beta}}}
  \left(
  \frac{\Gamma(\xi/2)}{\Gamma\left(\frac{1+\xi}{2}\right)}
  \right)^{2}
  \tan(\pi \xi/2)
  \,.
\end{equation}
We introduced the notation $n[z,\sigma]$ to indicate that the density
is a function of $z$ and the hard core diameter $\sigma$. Here
$\sigma=0$, but we shall use that notation later on.

Equation~(\ref{eq:density-sigma0}) together with the thermodynamic
relation $n=z \partial (\beta p)/\partial z$ leads to
\begin{equation}
  \label{eq:pressure-beta<2}
  \beta p= b_{\psi}(\beta) z^{\frac{4}{4-\beta}}
  \,.
\end{equation}
This is the exact pressure--fugacity relationship for point-like
particles when $\beta<2$. Notice that both~(\ref{eq:Fisher})
and~(\ref{eq:ansatzGT}) are compatible with~(\ref{eq:pressure-beta<2})
when $\sigma=0$ and $\beta<2$.

At $\beta=2$, the density diverges, as expected, due to the collapse
phenomenon.  As already noticed in Ref.~\cite{KS}, a naive analytic
continuation of~(\ref{eq:density-sigma0}) for $\beta\geq 2$ does not
give the correct density--fugacity relationship beyond the collapse:
the function $n[z,0]$ diverges at the thresholds $\beta=\beta_l$ and
can become negative, it cannot represent the density in the region
$2\leq\beta<4$.


\subsection{The Coulomb gas beyond $\beta=2$ in the low density limit}

The method~\cite{KS} to study the properties of the Coulomb gas for
$\beta>2$ and $\sigma\neq0$ when $n^2\sigma\ll 1$ is based on the
electroneutrality sum rule
\begin{equation}
  \label{eq:neutrality}
  n_{+}= \int_{\mathbb{R}^2} \left[ n^{(2)}_{+-}(r) - n^{(2)}_{++}(r)
  \right]\,d\r
\end{equation}
It is believed that the truncated correlation functions
$n^{(2)T}_{qq'}(r;z,\sigma)$ are well defined for $\sigma=0$ up to
$\beta<4$. Assuming that the difference
$n^{(2)T}_{qq'}(r;z,\sigma)-n^{(2)T}_{qq'}(r;z,0)$ is negligible for
$r>\sigma$, and using the electroneutrality sum
rule~(\ref{eq:neutrality}), Kalinay and \v{S}amaj~\cite{KS} propose
that the density $n[z,\sigma]$ for the Coulomb gas with hard core
$\sigma$ is given by
\begin{equation}
  \label{eq:n-howto}
  n[z,\sigma]=n[z,0]- 4\pi \int_0^{\sigma} \left[ n^{(2)}_{+-}(r;z,0)
  - n^{(2)}_{++}(r;z,0) \right]\,r\,dr
\end{equation}
in the limit $n\sigma^2\ll 1$.

Using the dominant order term in the small-$r$ expansion of the
correlation functions, $n_{+-}^{(2)}(r)-n_{++}^{(2)}(r) \sim z^2
r^{-\beta}$, Kalinay and \v{S}amaj~\cite{KS} obtained the first
correction
\begin{equation}
  n[z,\sigma]=n[z,0]-4\pi z^2 \frac{\sigma^{2-\beta}}{2-\beta}
\end{equation}
and they showed that it cancels the pole at $\beta=2$ from $n[z,0]$.


\section{The equation of state in the fugacity format}
\label{sec:pressure-general}

\subsection{The short distance expansion of correlation functions and the
  operator product expansion} 

To proceed further with the program proposed by Kalinay and
\v{S}amaj~\cite{KS}, we need to compute the higher order terms of the
short-distance expansion of the correlation functions. In a previous
work~\cite{GT-small-r}, we showed how the operator product expansion
for the exponential fields in the sine-Gordon model can be used to
obtain the short distance expansion of the correlation functions of
the Coulomb gas. Let us recall and extend some of the results from
Ref.~\cite{GT-small-r}.

The operator product expansion for the exponential fields of the
sine-Gordon models
reads~\cite{Fateev-Fradkin-Lukyanov-Zad2-descendent}
\begin{eqnarray}
  \label{eq:OPE}
  \langle e^{i b Q_1 \phi(0)} e^{i b Q_2 \phi(\r)} \rangle&=&
  \sum_{n=-\infty}^{n=+\infty} \Bigg[
  C_{Q_1 Q_2}^{n,0}(r) \langle e^{ib (Q_1+Q_2+n)\phi} \rangle \\
  &&+
  C_{Q_1 Q_2}^{n,2}(r) \langle (\partial\phi)^2 (\bar{\partial}\phi)^2
  e^{ib (Q_1+Q_2+n)\phi} \rangle
  +\cdots \Bigg]
  \nonumber
\end{eqnarray}
where the dots denote subdominant contributions from higher order
descendant fields
$\prod_{i}\partial^{m_i}\phi\prod_{j}\bar{\partial}^{n_j}\phi\ e^{ib
(Q_1+Q_2+n)\phi}$, where only the fields with
$\sum_{i}m_i=\sum_{j}n_j=m$ have non vanishing expectation
value. The level $m=1$ field has zero expectation
value because it is a total
derivative~\cite{Fateev-Fradkin-Lukyanov-Zad2-descendent}. 

The functions $C_{Q_1 Q_2}^{n,m}(r)$ of
the operator product expansion have the following
form~\cite{Fateev-Fradkin-Lukyanov-Zad2-descendent}
\begin{equation}
    \label{eq:OPE-coefs-C}
  C_{Q_1 Q_2}^{n,m}(r)=z^{|n|}
  r^{m+\beta Q_1 Q_2 + n \beta (Q_1+Q_2) + 2|n|(1-\frac{\beta}{4}) +n^2
    \beta /2} f_{Q_1 Q_2}^{n,m}(z^2 r^{4-\beta})
\end{equation}
where each $f_{Q_1 Q_2}^{n,m}$ admit a power series expansion of the form
\begin{equation}
  \label{eq:OPE-f}
  f_{Q_1 Q_2}^{n,0}(x)=\sum_{k=0}^{\infty} f_{k}^{n,m}(Q_1,Q_2) x^k
  \,.
\end{equation}
The connexion of the operator product expansion with the Coulomb gas
is clear by noticing that $\langle e^{ib(Q_1+Q_2+n)\phi} \rangle$ is
closely related to the excess chemical potential of an external charge
$Q_1+Q_2+n$ introduced in the plasma~\cite{Samaj-guest-charges}. Also,
the coefficients $f_{k}^{n,0}(Q_1,Q_2)$ are expressible in terms
$(n+2k)$-fold Coulomb type integrals: they are the partition functions
of a system with two fixed point charges $Q_1$ and $Q_2$ separated by
a distance $1$ and with $n$ (positive for $n>0$, negative for $n<0$)
point particles and $k$ pairs of positive and negative point particles
approaching the fixed particles. For explicit expressions of some of
these coefficients see
Refs.~\cite{Fateev-Fradkin-Lukyanov-Zad2-descendent,GT-small-r}.

A few technical details, explained in greater detail in
Ref.~\cite{GT-small-r}, should be kept in mind when using the
expansion~(\ref{eq:OPE}) to compute the correlation functions of the
Coulomb gas. First, the $(n+2k)$-fold Coulomb type integrals in the
coefficients $f_{k}^{n,m}(Q_1,Q_2)$ are defined for a certain range of
values of $Q_1$, $Q_2$ and $\beta$ since we are dealing with point
particles, in order to avoid the collapse. Beyond those ranges an
analytic continuation should be used. 

Second, for certain values of $Q_1$, $Q_2$ and $\beta$ different terms
in the expansion~(\ref{eq:OPE}) can become of the same order. When
this occurs the coefficient of each term usually has a pole, but
adding all contributions of the same power in $r$ and taking the
appropriate limit gives a finite result and logarithmic terms $\ln
(zr^{(4-\beta)/2})$ appear. One important case where this happens is
in the computation of $n_{+-}^{(2)}$. As clearly seen from
Eq.~(\ref{eq:OPE}), when $Q_1=-Q_2=1$, the terms for $n$ and $-n$ are
of the same order in $r$. One consequence is the appearance of
$\ln(zr^{(4-\beta)/2})$ terms in the expansion of $n_{+-}^{(2)}(r)$
coming from the contributions of terms $|n|\geq1$
in~(\ref{eq:OPE}). However, for $n=0$, the corresponding terms in
Eq.~(\ref{eq:OPE}) do not have a logarithmic correction. For details
see Ref.~\cite{GT-small-r}. This last remark is important for the
following analysis, and, as we shall see, it is the reason why the
first term of the equation of state~(\ref{eq:ansatzGT}) is an analytic
series in the fugacity.

Using~(\ref{eq:density-correl}) and the operator product
expansion~(\ref{eq:OPE}), including the contributions of all
descendant fields, we find that the general form of the short-distance
expansion of the correlation functions is
\begin{eqnarray}
  \label{eq:n++}
  n_{++}^{(2)}(r)&=&
  \sum_{n=-\infty}^{n=+\infty}
  \sum_{m=0}^{+\infty}
  \sum_{k=0}^{+\infty}
  n^{++}_{n,m,k} \frac{1}{r^4}
  \left(zr^{\frac{4-\beta}{2}}\right)^{\frac{\beta(2+n)^2+4m}{4-\beta}}
  \left(zr^{\frac{4-\beta}{2}}\right)^{2k+2+|n|}\\
  \label{eq:n+-}
  n_{+-}^{(2)}(r)&=&
  \sum_{n=-\infty}^{n=+\infty}
  \sum_{m=0}^{+\infty}
  \sum_{k=0}^{+\infty}
  n^{+-}_{n,m,k} \frac{1}{r^4}
  \left(zr^{\frac{4-\beta}{2}}\right)^{\frac{\beta n^2+4m}{4-\beta}}
  \left(zr^{\frac{4-\beta}{2}}\right)^{2k+2+|n|}
\end{eqnarray}
The indexes used are $m$ for the order of the descendant field
(remember that the term for $m=1$ is zero), $n$ for the number of
particles of sign $\text{sgn}(n)$ added and $k$ for the number of pair
of positive and negative particles added. The coefficients
$n^{+-}_{n,m,k}$ and $n^{++}_{n,m,k}$ depend on $\beta$ and eventually
on $\ln(zr^{(4-\beta)/2})$, except those for neutral configurations
$n^{+-}_{0,0,k}$ and $n^{++}_{-2,0,k}$ which depend only on $\beta$ as
explained above.


\subsection{The fugacity expansions of the density and the pressure}

Replacing~(\ref{eq:n++}) and~(\ref{eq:n+-}) into~(\ref{eq:n-howto})
leads to the following form for the density
\begin{equation}
  \label{eq:n-general}
  n[z,\sigma]=n[z,0]+
  \sum_{n=-\infty}^{+\infty} \sum_{m=0}^{+\infty} \sum_{k=0}^{+\infty}
  \frac{1}{\sigma^2}
  \left[
    c_{n,m,k}^{+-} \zeta^{\frac{4m+\beta n^2}{4-\beta}}+
    c_{n,m,k}^{++} \zeta^{\frac{4m+\beta (n+2)^2}{4-\beta}}
    \right]
  \zeta^{2k+2+|n|}
\end{equation}
with $n[z,0]$ given by Eqs.~(\ref{eq:density-sigma0})
and~(\ref{eq:bpsi}) and $\zeta=z \sigma^{(4-\beta)/2}$. Finally, using
$n=z \partial (\beta p)/\partial z$, the fugacity expansion of
pressure is of the form
\begin{eqnarray}
  \label{eq:betap-prelim}
  \beta p&=& b_{\psi}(\beta) z^{4/(4-\beta)} \\
  &+&
  \sum_{n=-\infty}^{\infty} \sum_{m=0}^{+\infty} \sum_{k=0}^{+\infty}
  \frac{1}{\sigma^2}
  \zeta^{\frac{4m+\beta n^2}{4-\beta}}
  \left[
    p^{+-}_{n,m,k} \zeta^{2k+2+|n|}
    +p^{++}_{n-2,m,k} \zeta^{2k+2+|n-2|}
    \right]
  \nonumber
\end{eqnarray}
Notice that the terms corresponding to $m=0$ and $n=0$ are analytic in
$z$. These, together with the terms $m=0$ and $n=1$, reproduce the
ansatz~(\ref{eq:Fisher}) from Fisher \textit{et
at.}~\cite{FLL}. Writing these terms apart in~(\ref{eq:betap-prelim})
the pressure--fugacity relationship can finally be written as
announced in the introduction
\begin{equation}
  \label{eq:betap-gen}
  \beta p=\frac{1}{\sigma^2}\sum_{l=1}^{\infty} \bar{b}_{2l}(\beta)
  \zeta^{2l} +b_{\psi}(\beta) z^{\frac{4}{4-\beta}}\left[
  1+e_{1,0}(\beta,\zeta)\right]
  +\frac{1}{\sigma^2}\sum_{m=0}^{\infty}\sum_{n}\nolimits^{*}
  \zeta^{\frac{4m+\beta n^2}{4-\beta}} \tilde{e}_{n,m}(\beta,\zeta)
  \,.
\end{equation}
The sum $\sum_{n}\nolimits^{*}$ does not contains the terms $m=0$ and
$n=0, 1$ since they are explicitly written apart: the sum is for
$n\geq 2$ when $m=0$ and for $n\geq 0$ for $m\geq 1$. 

The functions $\tilde{e}_{n,m}(\beta, \zeta)$ and $e_{1,0}(\beta,
\zeta)$ admit a power series expansion in terms of $\zeta$ and $\ln
\zeta$ given in Eqs.~(\ref{eq:tilde-e}) and~(\ref{eq:e}). They are at
least of order $\zeta^2=z^2 \sigma^{(4-\beta)}$, so they vanish in the
limit $n\sigma^2\to0$ for fixed $z$ and $\beta<4$: they are
irrelevant. On the other hand, in the analytic part of $\beta p$ as a
series in $\zeta$ (first sum in~(\ref{eq:betap-gen})), the $l$-th term
becomes relevant for $\beta>\beta_l=4[1-(1/(2l))]$, i.~e.~it diverges
in the limit $n\sigma^2\to0$.

At $\beta=\beta_l$, the term $z^{4/(4-\beta)} b_{\psi}(\beta)$
has a pole, as seen from~(\ref{eq:bpsi}), but it is expected that the
coefficient $\sigma^{-2}\bar{b}_{2l}(\beta) \zeta^{2l}$, which becomes
relevant at that point, also has a pole and cancels the divergence from
$z^{4/(4-\beta)} b_{\psi}(\beta)$. This was checked at
$\beta=\beta_1=2$ for the first term $l=1$ in Ref.~\cite{KS}. In the
next section we check that these cancellation also takes place at
$\beta=\beta_{2}=3$.


\section{Explicit calculations}
\label{sec:pressure-coefs}

\subsection{The equation of state in the fugacity format}

In Ref.~\cite{GT-small-r} we computed explicitly the short-distance
expansion of the correlation functions~(\ref{eq:n++})
and~(\ref{eq:n+-}) up to order $r^{8-3\beta}$, that is, we computed
the terms corresponding to $(n,m,k)=(-2,0,0)$ and $(n,m,k)=(-1,0,0)$
for $n_{++}^{(2)}$, and $(n,m,k)=(0,0,0)$, $(n,m,k)=(0,0,1)$ and
$(n,m,k)=(\pm 1,0,0)$ for $n_{+-}^{(2)}$. The neutral configurations:
for $n_{++}^{(2)}$, $(n,m,k)=(-2,0,0)$ gives the order $r^{4-2\beta}$,
and for $n_{+-}^{(2)}$, $(n,m,k)=(0,0,0)$ gives the order
$r^{-\beta}$, and $(n,m,k)=(0,0,1)$ the order $r^{4-2\beta}$. The
configurations with at most one charge $\pm1$ give the order
$r^{2-\beta}$ [$(n,m,k)=(-1,0,0)$ for $n_{++}^{(2)}$ and $(n,m,k)=(\pm
1,0,0)$ for $n_{+-}^{(2)}$]. Explicitly,
\begin{eqnarray}
  \label{eq:n+--expl}
  n_{+-}^{(2)}(r)&=&
  z^2 r^{-\beta}
  +z^3 \langle e^{ib\phi} \rangle r^{2-\beta}
  \left( \tilde{n}^{+-}_{3} -\pi\beta^2 \ln \left[\left(\frac{\pi
  z}{\gamma(\beta/4)}\right)^{\frac{2}{4-\beta}}r\right]\right)\\
  &&+ z^4 r^{4-2\beta} \tilde{n}^{+-}_{4} 
  +O(r^4,r^{8-3\beta},r^{6-2\beta})
  \nonumber\\
  \label{eq:n++-expl}
  n_{++}^{(2)}(r)&=&
  z^3 \langle e^{ib\phi} \rangle r^{2-\beta} \tilde{n}_{3}^{++}
  + z^4 r^{4-2\beta} \tilde{n}_{4}^{++} 
  +O(r^4,r^{8-3\beta},r^{6-2\beta})
\end{eqnarray}
with $\langle e^{ib\phi} \rangle$ given by Eq.~(\ref{eq:eibphi}), and
\begin{eqnarray}
  \label{eq:use-Iprimeb}
  \tilde{n}_{3}^{+-}&=&
  -\frac{\pi\beta^2}{4}
  \Bigg[\frac{4}{\beta}I'_b(1) - 4 + 4 C 
    \\
    &&+ \psi(\frac{\beta}{2})
    +\psi(-\frac{\beta}{2}) + \psi(1-\frac{\beta}{2}) + \psi(1+\frac{\beta}{2})
    \Bigg]
  \nonumber\\
  \tilde{n}_{4}^{+-}&=&J(\beta,-\beta,\beta)
  \label{eq:use-J}
  \\
  \tilde{n}_{3}^{++}&=& \pi \gamma(1-\frac{\beta}{2})^2
  \gamma(\beta-1)\\
  \tilde{n}_{4}^{++}&=&
  -\frac{4\pi^2}{(2-\beta)^2}
  \gamma(1-\frac{\beta}{4})^3\gamma(-1+\frac{3\beta}{4})
\end{eqnarray}
where $\psi(x)=d\ln \Gamma(x)/dx$ is the digamma function and $C=-\psi(1)$ is
the Euler constant. The functions $I_b'(1)=\partial I_b(Q)/\partial
Q|_{Q=1}$ and $J(\beta,-\beta,\beta)$ are detailed in the appendix.

The use of~(\ref{eq:n-howto}) leads to the density
\begin{eqnarray}
  \label{eq:n-z4}
  n[z,\sigma]&=&
  z\langle e^{ib\phi} \rangle
  \left\{
  2-\frac{4\pi z^2\sigma^{4-\beta}}{4-\beta}
  \left(
  \tilde{n}_{3}^{+-}-\tilde{n}_{3}^{++}
  -\frac{\pi\beta^2}{4-\beta}
  \left[
    \ln \left[\left(\frac{\pi
	z}{\gamma(\beta/4)}\right)^{2}\sigma^{4-\beta}\right]
    -1
    \right]
  \right)
  \right\}
  \nonumber\\  &&
  -\frac{4\pi z^2\sigma^{2-\beta}}{2-\beta}
  -\frac{2\pi z^4\sigma^{2(3-\beta)}}{3-\beta}
  (\tilde{n}_{4}^{+-}-\tilde{n}_{4}^{++})
  +O(\sigma^6,\sigma^{10-3\beta},\sigma^{8-2\beta})
\end{eqnarray}

Integrating the thermodynamic relation $n=z\partial(\beta p)/\partial
z$, we find the equation of state in the fugacity format
\begin{equation}
  \label{eq:betap-z4}
  \beta p =
  b_{\psi}(\beta) z^{4/(4-\beta)}
  \left[
  1+e(\beta,\zeta)
  \right]+
  \frac{1}{\sigma^2}
  \left[
    \frac{-2\pi}{2-\beta} \zeta^2
    +\bar{b}_4(\beta) \zeta^4
    +O(\zeta^6)
    \right]
\end{equation}
with
\begin{equation}
  \label{eq:b4-def}
  \bar{b}_{4}(\beta)=
  -\frac{\pi}{2(3-\beta)}
  \left[
    J(\beta,-\beta,\beta)+\frac{4\pi^2}{(2-\beta)^2}
      \gamma(1-\frac{\beta}{4})^3\gamma(-1+\frac{3\beta}{4})
    \right]
\end{equation}
and
\begin{eqnarray}
  e(\beta,\zeta)&=&
  -\frac{4\pi \zeta^2}{(4-\beta)^2(6-\beta)}
  \times
  \nonumber
  \\&&
  \Bigg\{
  (4-\beta)
  \Bigg(
  \frac{-\pi\beta^2}{4}
  \left[
    \frac{4}{\beta}I'_b(1)-4+4C
    + \psi(\frac{\beta}{2})
    +\psi(-\frac{\beta}{2}) + \psi(1-\frac{\beta}{2}) + \psi(1+\frac{\beta}{2})
    \right]
  \nonumber\\
  &&
  -\pi\gamma(1-\frac{\beta}{2})^2\gamma(\beta-1)
  \Bigg)
  -\pi\beta^2
  \left[
    \frac{2(\beta-5)}{6-\beta}
    +\ln
      \left(\frac{\pi\zeta}{\gamma(\beta/4)}\right)^2
    \right]
  \Bigg\}
  +o(\zeta^2)
  \label{eq:e-beta-zeta}
\end{eqnarray}
We confirm that the function $e(\beta, z\sigma^{(4-\beta)/2})\to 0$
when $n\sigma^2\to0$ at fixed $z$.


\subsection{Cancellations at $\beta=3$ for the relevant terms}
\label{sec:cancels-beta=3}

At $\beta=3$, the contribution to $\beta p$ from $n[z,0]$ is
\begin{equation}
  b_{\psi}(\beta) z^{4/(4-\beta)}\simat{\beta\to3}
  \frac{\pi^3}{8}
  \gamma(1/4)^4
  z^{4}
  \frac{1}{\beta-3}
\end{equation}
On the other hand, in the appendix it is shown that
\begin{equation}
  \label{eq:b4-beta=3}
  \bar{b}_{4}(\beta)\simat{\beta\to3}
  -\frac{\pi^3}{8}  \gamma(1/4)^4
  \frac{1}{\beta-3}
  \,.
\end{equation}
Thus the divergences of each term at $\beta=3$ cancel each other
yielding a finite result for the pressure. The cancellation mechanism
conjectured in Ref.~\cite{KS}, indeed take place at $\beta=3$.

\subsection{Cancellations at $\beta=2$ for irrelevant terms}

Actually a more complex mechanism of cancellations of divergences
appears to take place, at $\beta=\beta_l$, when the generically
nonanalytic contributions in $z^{4/(4-\beta)}$ become analytic
$z^{2l}$, also with the irrelevant terms (terms that vanish when
$n\sigma^2\to0$). To illustrate this, notice that at $\beta=2$,
$\bar{b}_4(\beta) \zeta^{4}/\sigma^2$ has a pole,
but so does the term $b_{\psi}(\beta) z^{4/(4-\beta)} e(\beta,\zeta)$
which is also of order $\zeta^4$ at $\beta=2$. The finiteness of the
correlation functions in the region up to $\beta<4$ (in particular at
$\beta=2$ for the present case) ensures that both divergent
contributions cancel each other.

The calculations of the short distance expansion of the
correlation functions at $\beta=2$ has been done in
Ref.~\cite{GT-small-r}. Using the results from Ref.~\cite{GT-small-r},
in particular the results from Appendices A and B of
Ref.~\cite{GT-small-r}, it is easy to check that the contribution of
terms of order $\zeta^4$ in the pressure are finite:
\begin{eqnarray}
\beta p&\eqat{\beta=2}&
-\pi z^2 \left[-1+2C+2\ln(\pi z\sigma)\right]
-\frac{\pi^3 z^4 \sigma^2}{4}
\big\{
-3-4(C+\ln \pi)[1+2(C+\ln\pi)]
\nonumber\\
&&+4\left[-1+2\ln (z\sigma)+4(\ln\pi+C)\right]
\ln(z\sigma)
\big\}
+o(z^4\sigma^2)
\end{eqnarray}
We have also written the relevant contribution (nonvanishing when
$\sigma\to0$) which was computed in Ref.~\cite{KS}.

If the conjecture that the truncated correlation functions are finite
up to $\beta=4$, this cancellation mechanism of irrelevant terms
should take place at other values of $\beta$ where the coefficients in
Eq.~(\ref{eq:betap-gen}) diverge. For instance, in~(\ref{eq:betap-z4})
the product $b_{\psi}(\beta)e(\beta,\zeta)$ has a pole at
$\beta=3$. At this value of $\beta$, the nonanalytic contribution to
the pressure $b_{\psi}(\beta)e(\beta,\zeta) z^{4/(4-\beta)}$ becomes
analytic of order $\zeta^{6}$. The pole at $\beta=3$ of
$b_{\psi}(\beta)e(\beta,\zeta)$ should be canceled with a similar
diverging term from $\bar{b}_6(\beta) \zeta^6/\sigma^2$, which we have
not computed here.


\section{Comparison with Monte Carlo simulations}
\label{sec:mc}

In the canonical format, the excess dimensionless free energy per
particle is given by
\begin{equation}
  f(n,\beta)=\frac{-\beta p}{n}+\ln \zeta
\end{equation}
where $\zeta$ should be expressed in terms of the density $n$ by
inverting the relation~(\ref{eq:n-general}). The excess internal
energy per particle $u^{\text{exc}}$ and the excess specific heat at
constant volume per particle $c_{V}^{\text{exc}}$ can be obtained
as
\begin{equation}
  u^{\text{exc}}=\frac{\partial f(n,\beta)}{\partial \beta}
\end{equation}
and
\begin{equation}
  c_{V}^{\text{exc}}=-\beta^2 \frac{\partial^2 f(n,\beta)}{\partial \beta^2}
  \,.
\end{equation}
Using the expressions~(\ref{eq:n-z4}) and~(\ref{eq:betap-z4}) for the
density and the pressure, accurate to order $O(\zeta^6)$, obtained in
the previous section, we numerically inverted the density--fugacity
relationship~(\ref{eq:n-z4}) and computed the internal energy and
specific heat for two low density packing fractions $\eta=\pi n
\sigma^2/4$, $\eta=5\cdot10^{-4}$ and $\eta=5\cdot10^{-3}$. In
Ref.~\cite{Caillol-Levesque}, Monte Carlo simulations of the
two-dimensional Coulomb gas where performed for these two packing
fraction values.

%
%
\begin{figure}
  \includegraphics[width=\GraphicsWidth]{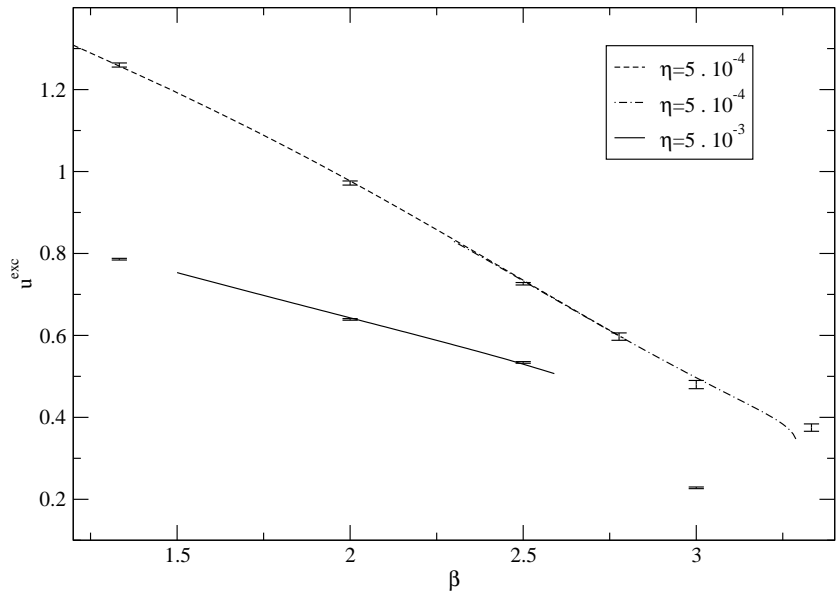}
  \caption{
    \label{fig:uexc}
    The internal energy $u^{\text{exc}}$ as a function of $\beta$. The
    lines are our analytical results and the bars are the Monte Carlo
    simulation results. For $\eta=5\cdot 10^{-4}$ the dashed line
    shows the results obtained with the full
    expression~(\ref{eq:betap-z4}), while the dash-dot line shows the
    results obtained from~(\ref{eq:betap-z4-nd}).
}
\end{figure}
%
%
%

\begin{figure}
  \includegraphics[width=\GraphicsWidth]{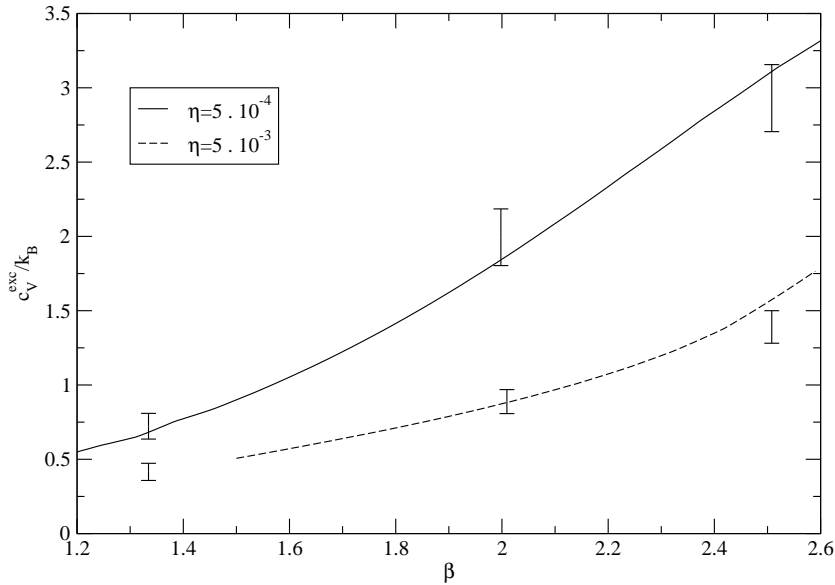}
  \caption{
    \label{fig:cvexc}
    The specific heat $c_{V}^{\text{exc}}$ as a function of
    $\beta$. The lines are our analytical results and the bars are the
    Monte Carlo simulation results.  }
\end{figure}

Our analytical results are compared to the Monte Carlo simulation ones
in figures~\ref{fig:uexc} and~\ref{fig:cvexc}. In
figure~\ref{fig:uexc}, we plot the internal energy $u^{\text{exc}}$ as
a function of the inverse temperature $\beta$. For $\eta=5\cdot
10^{-4}$, we show two curves. The dashed line corresponds to the
results obtained from Eq.~(\ref{eq:betap-z4}). As it was discussed in
the preceding section, the term $b_{\psi}(\beta)e(\beta,\zeta)$ from
Eq.~(\ref{eq:betap-z4}) has a pole at $\beta=3$, which should be
canceled with the next order term
$\bar{b}_{6}(\beta)\zeta^6/\sigma^2$ which has not been
computed. For this reason, the comparison with Monte Carlo results can
only be done for $\beta<3$. However, since this term vanishes when
$\eta\to 0$, we decided to compare the Monte Carlo results with those
obtained from our analytical formulas by omitting this ``irrelevant''
term. This is shown in figure~\ref{fig:uexc} with the dot-dash
line. Explicitly, this last curve is obtained by approximating the
pressure by
\begin{equation}
  \label{eq:betap-z4-nd}
  \beta p =
  b_{\psi}(\beta) z^{4/(4-\beta)}
  +
  \frac{1}{\sigma^2}
  \left[
    \frac{-2\pi}{2-\beta} \zeta^2
    +\bar{b}_4(\beta) \zeta^4
    \right]
\end{equation}
instead of using Eq.~(\ref{eq:betap-z4}). Eq.~(\ref{eq:betap-z4-nd})
can be used up to the next pole at $\beta=\beta_3=10/3$. For very low
volume fractions, $\eta=5\cdot10^{-4}$, the agreement with Monte Carlo
simulations is very good, both using the correct
formula~(\ref{eq:betap-z4}), for $\beta<3$, or the ``truncated''
one~(\ref{eq:betap-z4-nd}), for $2<\beta<10/3$. For higher volume
fractions, $\eta=5\cdot10^{-3}$, the agreement is still very good when
using the complete formula~(\ref{eq:betap-z4}). On the other hand, the
``truncated'' formula Eq.~(\ref{eq:betap-z4-nd}), does not give a good
agreement (curve not shown) as it can expected because the omitted
term becomes important at high volume fractions.

In figure~\ref{fig:cvexc}, we plot the specific heat
$c_{V}^{\text{exc}}$ as a function of $\beta$. The agreement with
Monte Carlo simulations is fairly good, even at the relatively high
volume fraction $\eta=5\cdot 10^{-3}$. In any case, the agreement with
the simulations is much better than the one obtained with only the
first order term $\bar{b}_{2}(\beta)\zeta^2/\sigma^2$ in the pressure,
shown in Fig.~3 of Ref.~\cite{KS}, as expected.

%
%
%
%
%


\section{Conclusion}

Using the exact results for the thermodynamics of the two-dimensional
Coulomb gas of point particles for $\beta<2$~\cite{ST}, the
short-distance expansion of the density correlations
functions~\cite{GT-small-r}, and the program proposed by Kalinay and
\v{S}amaj~\cite{KS}, we have derived the general form of the equation
of state in the fugacity format~(\ref{eq:ansatzGT}), for the
two-dimensional Coulomb gas composed of small core diameter $\sigma$
particles and for $\beta<4$. We explicitly computed the second
corrections due to the hard core, up to terms $O(\zeta^6)$ in the
activity $\zeta$, the first corrections (order $\zeta^2$) where
computed in Ref.~\cite{KS}.

The general form of the equation of state~(\ref{eq:ansatzGT}) is
compatible with the fact that an analytic expansion of the pressure in
powers of the fugacity have finite Mayer coefficients up to order $2l$
for $\beta>\beta_l$, confirming the findings of Gallavotti and
Nicol\'o~\cite{GN}. However, the explicit calculations we performed
show that the pressure does not have any singularities at
$\beta=\beta_1=2$ nor at $\beta=\beta_2=3$, contrary to the conjecture
of a series of intermediate phase transitions proposed by Gallavotti
and Nicol\'o~\cite{GN}, and supporting the arguments of Fisher
\textit{et at.}~\cite{FLL} against this conjecture. But the general
form for the equation of state~(\ref{eq:ansatzGT}) we found is more
complex than the ansatz~(\ref{eq:Fisher}) proposed by Fisher
\textit{et at.}~\cite{FLL}. Finally, we compared our results against
Monte Carlo simulations results and we found good agreement.


\section*{Acknowledgments}

The author thanks L.~\v{S}amaj for valuable discussions. This work was
supported by a ECOS Nord/COLCIENCIAS action of French and Colombian
cooperation.


\begin{appendix}

\def\theequation{\Alph{section}.\arabic{equation}}

\section{Appendix: Technical details}

The function $I'_b(1)$ appearing in Eq.~(\ref{eq:use-Iprimeb}) is
defined as $I'_b(1)=\partial I_b(Q)/\partial Q|_{Q=1}$ with $I_b(Q)$
given by Eq~(\ref{eq:IbQ}). Explicitly,
\begin{equation}
  \label{eq:Iprimeb}
    I'_b(1)=
  \frac{\beta}{4} \int_0^{\infty}
  \frac{dt}{t}\,
  \left[
    -4  e^{-2t}
    +\frac{t \sinh(\beta t)}{\sinh t \cosh[(1-\frac{\beta}{4})t]
      \sinh(\beta t/4)}
    \right]
  \,.
\end{equation}
This expression converges only for $\beta<2$. To use it beyond
$\beta>2$ it is useful to write it as
\begin{equation}
  \label{eq:Iprimeb-descompo}
    \frac{4}{\beta}I'_b(1)=\I_1(\beta)+\I_2(\beta)
\end{equation}
where
\begin{eqnarray}
  \I_1(\beta)&=&2\int_0^{\infty}
  \left(-\frac{e^{-2t}}{t}+\frac{\cosh(\beta t/4)}{\sinh t
  \cosh[(1-\frac{\beta}{4})t]}\right)\,dt\\
  \I_2(\beta)&=&2\int_{0}^{\infty}\left(
  -\frac{e^{-2t}}{t}+\frac{\cosh(3\beta t/4)}{\sinh t
  \cosh[(1-\frac{\beta}{4})t]}\right)\,dt
\end{eqnarray}
The first integral, $\I_1(\beta)$, is well defined for $\beta<4$. The
second one, $\I_2(\beta)$, is defined for $\beta<2$, but it can be
analytically continued up to $\beta<10/3$, by writing it as
\begin{eqnarray}
  \I_2(\beta)&=&
  \frac{4}{2-\beta}+\frac{8}{3\beta-8}+\frac{2}{3-\beta}+\frac{8}{5\beta-16}
  \\
  &+&2\int_0^{\infty}\left[
    -\frac{e^{-2t}}{t}
    +2\frac{e^{-t(4+\beta)/2}-e^{-3t(4-\beta)}+e^{-t(4-\beta)}
      +e^{-t(10-3\beta)}}{(1-e^{-2t})(1+e^{-2t(1-\frac{\beta}{4})})}
    \right]\,dt.
  \nonumber
\end{eqnarray}


The function $J(\beta,-\beta,\beta)$ appearing in Eq.~(\ref{eq:use-J})
is a special case of the integral
\begin{equation}
    \label{eq:Dotsenko-integral-J-part}
  J(\beta Q,-\beta Q,\beta)=
  \int d^2 x \,d^2 y\,
  \frac{|x|^{\beta Q} |1-y|^{\beta Q}}{|y|^{\beta Q}
  |1-x|^{\beta Q} |x-y|^{\beta}}
  \,.
\end{equation}
for $Q=1$. Using the results from Ref.~\cite{Dotsenko-Picco-Pujol}, in
the appendix B of Ref.~\cite{GT-small-r} we showed that
\begin{equation}
  \label{eq:J}
  J(\beta,-\beta,\beta)=
  \left[
    s(\beta/2) J_{1}^{+}+s(\beta)J_{2}^{+}
    \right]^2+
  \left[
    s(\beta/2) J_{2}^{+}
    \right]^2
\end{equation}
with
\begin{eqnarray}
  J_{1}^{+}&=&
  \frac{\Gamma(1-\frac{\beta}{2})^3\Gamma(2-\frac{\beta}{2})}{
    \Gamma(2-\beta)\Gamma(3-\beta)}\,
  \F32 \left(1-\frac{\beta}{2},2-\frac{\beta}{2},-\frac{\beta}{2};
  2-\beta,3-\beta;1\right)\\
  J_{2}^{+}&=&
  \frac{\Gamma(1-\frac{\beta}{2})\Gamma(2-\frac{\beta}{2})
    \Gamma(1+\frac{\beta}{2})^2
  }{2}\,
  \F32 \left(2-\frac{\beta}{2},1+\frac{\beta}{2},\frac{\beta}{2};
  2,3;1\right)
  .
  \label{eq:J-end}
\end{eqnarray}
where we used the notation $s(x)=\sin (\pi x)$ and where
$\F32(a_1,a_2,a_3;b_1,b_2;z)=\sum_{k=0}^{\infty} \frac{(a_1)_k(a_2)_k
(a_3)_k}{(b_1)_k (b_2)_k k!}z^k$ is a generalized hypergeometric
function, and $(a)_k=\Gamma(a+k)/\Gamma(a)$ is the Pochhammer symbol.

In the appendix B of Ref.~\cite{GT-small-r} we proved that near
$\beta=2$,
\begin{equation}
  \label{eq:J-beta=2}
  J(\beta,-\beta,\beta)
  \eqat{\beta\to2}
  (2\pi)^2 \left[
    \frac{1}{(\beta-2)^2}+\frac{1}{\beta-2}
    +1
    \right]
  +O(\beta-2)
  \,.
\end{equation}

In order to verify the cancellation mechanism at $\beta=3$ presented
in Sec.~\ref{sec:cancels-beta=3}, we need the value of
$J(\beta,-\beta,\beta)$ at $\beta=3$.

For this purpose, it is convenient to write $J_{1}^{+}$ as
\begin{equation}
  J_{1}^{+}=\frac{
    \Gamma(1-\frac{\beta}{2})^3\Gamma(2-\frac{\beta}{2})}{\Gamma(4-\beta)^2} 
  \left[
    (2-\beta)(3-\beta)^2+(1-\frac{\beta}{2})(2-\frac{\beta}{2})
    (-\beta/2)(3-\beta)
    +S(\beta)
    \right]
\end{equation}
where
\begin{equation}
  S(\beta)=\sum_{k=2}^{+\infty}
  \frac{
    (1-\frac{\beta}{2})_k (2-\frac{\beta}{2})_k (-\beta/2)_k}{
    (4-\beta)_{k-1} (4-\beta)_{k-2} k! }
  \,.
\end{equation}
At $\beta=3$, we have $S(3)=-\gamma(1/4)^2/(16\pi)$. Then at $\beta=3$,
\begin{equation}
  J_{1}^{+}\eqat{\beta=3}
  \frac{\pi}{2}\gamma(1/4)^2
  \,.
\end{equation}
On the other hand $J_{2}^{+}$ can be evaluated directly at $\beta=3$,
using~\cite{Mathematica} 
\begin{equation}
\F32(1/2,5/2,3/2;2,3;1)=\frac{8}{9\pi}\gamma(1/4)^2
\,.
\end{equation}
We find that $J_{2}^{+}=-J_{1}^{+}$ at $\beta=3$. This yields
\begin{equation}
  J(3,-3,3)=\frac{\pi^2}{2}\gamma(1/4)^4 \,.
\end{equation}
This result, combined with the explicit expression~(\ref{eq:b4-def}) for
$\bar{b}_{4}(\beta)$, gives the behavior~(\ref{eq:b4-beta=3}) for
$\bar{b}_4(\beta)$ near $\beta=3$.

\end{appendix}


\end{document}